\documentclass[twocolumn,showpacs,preprintnumbers,amsmath,amssymb]{revtex4}

\usepackage{graphicx}
\usepackage{dcolumn}
\usepackage{bm}

\newcommand{\eqb}{\begin{equation}}
\newcommand{\eqe}{\end{equation}}
\newcommand{\pd}{\partial}

\newcommand{\eab}{\begin{eqnarray}}
\newcommand{\eae}{\end{eqnarray}}
\newcommand{\ra}{\right\rangle}
\newcommand{\la}{\left\langle}
\newcommand{\e}{\mbox{e}}

\newcommand{\La}{\Lambda}

\begin{document}
\title{Spontaneously broken, local Z$_2$ symmetry and the anatomy of a lepton}
\author{Ralf Hofmann}\vspace{0.3cm}
\affiliation{Institut f\"ur Theoretische Physik,
Universit\"at Heidelberg,
Philosophenweg 16,
69120~Heidelberg,
Germany}

\begin{abstract}
In the framework of the gauge symmetry 
$\mbox{SU(2)}_{\tiny{\mbox{CMB}}}\times \mbox{SU(2)}_{\tiny{e}}\times \mbox{SU(2)}_{\tiny{\mu}}\times 
\mbox{SU(2)}_{\tiny{\tau}}$ we provide a microscopic argument why a 
spontaneously broken, local Z$_2$ symmetry surviving the 1$^{\tiny\mbox{st}}$ order 
phase transition to the confining phase of one of the SU(2) 
factors generates two stable spin-1/2 excitations of vastly different mass. 
A mixture of Fermi-Dirac and Bose-Einstein 
distributions with the former favored by a factor two over the latter is obtained 
by applying inequivalent, local-in-Euclidean-time Z$_2$ transformations 
to one and the same field configuration contained in the description 
of the center-vortex condensate. 


\end{abstract} 

\maketitle

{\sl Introduction.} In \cite{PRL2} we have constructed and applied a thermal theory for 
charged leptons. The underlying gauge symmetry for this theory is:
\eqb
\label{gs}
\mbox{SU(2)}_{\tiny{\mbox{CMB}}}\times \mbox{SU(2)}_{\tiny{e}}\times \mbox{SU(2)}_{\tiny{\mu}}\times 
\mbox{SU(2)}_{\tiny{\tau}}\,.
\eqe
At the present temperature of the Universe 
$T_{\tiny\mbox{CMB}}\sim 2.728\,K\sim 2.2\times10^{-4}$\,eV the dynamics due to the latter three 
gauge factors is confining. Moreover, the factor $SU(2)_{\tiny{\tau}}$ implies gauge dynamics which is
contaminated by strong interactions. As was noticed in \cite{PRL2}, the experimental observation 
that the CMB photon is very close to massless \cite{Williams1971} 
\footnote{The mass bound from the measurement of the 
Coulomb potential is $\sim 10^{-14}\,$eV.} implies that the theory 
associated with the gauge factor $SU(2)_{\tiny\mbox{CMB}}$ is at the magnetic 
side of the phase boundary 
to the electric phase. 
The gauge dynamics due to 
SU(2)$_{\tiny{\mbox{CMB}}}$ apparently is the only nonconfining dynamics 
in our present universe. The purpose of this Letter 
is to show that the spontaneously broken, local $Z_2$ symmetries associated with each of 
the factors in SU(2)$_{\tiny{e}}\times$SU(2)$_{\tiny{\mu}}\times$SU(2)$_{\tiny{\tau}}$ 
in their confining
phases render the respective two distinct stable excitations effectively 
spin-1/2 fermions. We provide statistical and microscopic arguments in favor for 
this claim. 

{\sl 't Hooft loop.} The confining 
phase of a pure SU(N) gauge theory is related to a vanishing expectation 
of the Polyakov loop. As was explained in \cite{Smilga1994,Korthales-Altes1999} 
and stressed in \cite{deForcrand2001} there are severe problems associated with this 
operator: it creates a static and
fundamental color source which is not in the physical Hilbert space, it is not defined 
at zero temperature, and it is plaqued by ultraviolet 
divergences in the continuum limit. Another, much healthier order parameter 
for the deconfinement-confinement 
transition is the vacuum expectation of the 't Hooft-loop operator 
$B[C]$ \cite{tHooft1978}. This operator is 
associated with a closed curve $C$ which is linked to another curve $C^\prime$ 
$n$ times. Both curves $C$ and $C^\prime$ are purely spatial when working in 
Minkowski signature. In 4D Euclidean space, $C$ and $C^\prime$ can also extend 
into the time direction. The action of the operator $B[C]$ 
results in a multi-valued gauge transformation 
$\Omega[C]$ of the field $A_i$ along the curve $C^\prime$. 
Namely, parametrizing $C^\prime$ by an angle 
$0\le\theta\le 2\pi$ we have 
\eqb
\label{tHooft}
\Omega[C](2\pi)=\Omega[C](0)\e^{\pm 2\pi i n/N}\,,
\eqe
where the +(-) sign refers to the clockwise (anticlockwise) sense of running through 
the curve $C^\prime$ \cite{tHooft1978}. In the absence of any preferred 
direction in the system (for the case in Eq.\,(\ref{gs}) this could be a homogeneous, 
external magnetic field with respect to the
unbroken U(1)$_{\tiny\mbox{SM}}$ gauge theory in the Standard Model \cite{PRL2}) 
the clockwise and anticlockwise sense are 
physically equivalent. $B[C]$ creates an elementary magnetic 
flux along the curve $C$ out of the ground state. In contrast to the Polyakov loop 
a nonvanishing expectation 
value of the 't Hooft-loop operator $B[C]$ indicates a spontaneous breakdown of a the (local) 
center symmetry $Z_{\tiny\mbox{N}}$ of SU(N) \cite{tHooft1978}. 
Such a phase is identified with the confining phase of the 
SU(N) Yang-Mills theory \cite{tHooft1978}. It is striking 
that the author of \cite{tHooft1978} stresses a similarity with 
fermionic Green functions in his discussion of correlation functions 
of 't Hooft-loop operators in Euclidean spacetime. This similarity is related to 
an ambiguity in the definition of the phase factor. 
  
The fields 
$\Phi_n(x)$ introduced in \cite{PRL1} 
are defined by the action of the 't Hooft loop. Explicitly, we have 
\eqb
\label{cvf}
\Phi_n(x)=\la \mbox{tr}\,{\cal P} \exp\left[ie\oint_{C^\prime_n(x)} dz^i A_i(z)\right]\ra \,,
\eqe
where $e$ denotes the Yang-Mills coupling constant, and ${\cal P}$ is the path-ordering symbol. 
In Eq.\,(\ref{cvf}) $C^\prime_n(x)$ denotes a spatial curve centered at $
x$ which is linked $n$ times to a curve $C$. The `radius' of $C^\prime_n(x)$ 
is not much larger than the core-size of the $n^{\tiny\mbox{th}}$ 
center vortex ($n=1,\cdots,\mbox{N}-1$) piercing it along the curve $C$
\footnote{The availability of such an $n^{\tiny\mbox{th}}$ 
center vortex is guaranteed in a {\sl condensate} of center vortices existing in a phase of 
spontaneously broken Z$_{\tiny\mbox{N}}$ symmetry \cite{PRL1}. This condensate is created 
by a diverging magnetic gauge coupling 
$g$ in the magnetic phase of the theory \cite{PRL1}.}. 

{\sl Statistical situation.} In \cite{PRL1} we have constructed the 
effective potential for the local 
field $\Phi_n$ from the requirements that (a) a matching 
to the magnetic phase takes place in thermal equilibrium and (b) 
the (spontaneously broken) Z$_{\tiny\mbox{N}}$ symmetry 
is implemented in the potential in a nontrivial way. Requirement (a) implies, 
that the Euclidean time dependence of the center-vortex 
condensate $\Phi_n$ is Bogomoln'yi-Prasad-Sommerfield (BPS) 
saturated along the compactified Euclidean time 
coordinate. Requirement (b) implies that the 
vacuum pressure vanishes identically in the 
confining phase. According to the proposed potential \cite{PRL1}
\eqb
\label{pot}
V_C^{(n)}\equiv\bar{v}_C^{(n)}v_C^{(n)}\,,
\eqe
where
\eqb
\label{sqrpot}
v_C^{(n)}\equiv i(\La_C^3/\Phi_{n}-\Phi_{n}^{\tiny{\mbox{N-1}}}/\La_C^{\tiny\mbox{N}-3})\,,
\eqe
requirements (a) and (b) are simultaneously 
only then satisfied if $\mbox{N}\to\infty$. For finite N and at the phase boundary 
the form of the BPS saturated solution along the compactified Euclidean time coordinate 
subject to the potential in Eq.\,(\ref{sqrpot}) indicates the breakdown of thermal 
equilibrium by itself \cite{Hofmann00}.  

Let us from now on only consider the case SU(2). This case is very likely 
relevant for an understanding of the statistics and the anatomy of 
charged leptons and neutrinos, see below and
\cite{Hofmann2004}. For our statistical consideration 
we assume that reheating process during the phase transition from the magnetic 
to the confining phase of the SU(2) gauge theory is efficient enough so that 
the system can be described thermodynamically not too long after the onset 
of vortex condensation. A check of this assumption could be done by using methods of 
nonequilibrium field theory subject to the potential of Eq.\,(\ref{pot}). It is mandatory 
for our argument that such a check will be performed in the future. 
For now we {\sl assume} a Euclidean field theory 
with compactified time coordinate $0\le\tau\le 1/T$ and the following action 
\cite{PRL1}
\eqb
\label{actN=2}
S_C=\int_0^{1/T} d\tau\int d^3x\left\{\left(\pd_\tau\Phi\right)^2+
\Lambda_C^2\left(\Lambda_C^2/\Phi-\Phi\right)^2\right\}\,.
\eqe
At the phase boundary the field $\Phi$ still satisfies periodic 
boundary conditions \footnote{Referring to the field $\Phi$ in our statistical discussion 
we do not mean the expectation as in Eq.\,(\ref{cvf}) 
but a field configuration as it contributes 
to the partition function.}. 
Due to the breakdown of thermal equilibrium the expectation of 
$\Phi$ for a short period (probably a fraction of the 
inverse Yang-Mills scale $\Lambda_C^{-1}$) 
develops a violent time dependence. Thermal equilibirum is restored as this expectation 
relaxes towards the minimum of the potential in (\ref{actN=2}). 
Since the spontaneously broken Z$_2$ symmetry is discrete the action 
of a genuinely local Z$_2$ transformation leads to a physical effect. 
Let us consider the following, local Z$_2$ transformations
\eqb
\label{z2trafo}
T_{\pm}=\mp2\Theta(\frac{1}{2T}-\tau)\pm 1\,
\eqe
where $\Theta$ denotes the usual Heaviside function ($\Theta(0)=1/2$). 
The transformations $T_{\mp}$ are depicted in Fig.\,1(a) and Fig.\,1(b), respectively. 
Applying $T_{\pm}$ to a field 
configuration $\Phi$ subject to $\Phi(0,\vec{x})=\Phi(1/T,\vec{x})$ 
generates boundary conditions $\Phi(0,\vec{x})=-\Phi(1/T,\vec{x})$ in two inequivalent 
ways \footnote{This can be seen by adding to the action density in 
(\ref{actN=2}) a small perturbation $B\Phi$ which breaks the Z$_2$ symmetry
explicitly.}. Both transformations change the action formally by a piece
\eqb
\label{change}
\Delta S_C=4\,\delta(0)\int d^3x\,(\Phi(1/(2T),\vec{x}))^2\,.
\eqe
$T_{\pm}$ creates and destroys an explicit center vortex of infinite spatial
extension, respectively. At $\tau=1/(2T)$ a 't Hooft-loop operator, 
defined by a circular loop say, in the 
$\tau-x_1$ plane and centered at $(\tau,\vec{x})$, 
creates or destroys one unit of magnetic flux, say, 
along the $x_2$ direction. The tranformations $T_{\pm}$ are the basic building blocks 
for all nontrivial, local Z$_2$ transformations 
along the Euclidean time interval which change periodic into antiperiodic boundary conditions. 
Global $Z_2$ transformations neither change the action (\ref{actN=2}) nor the nature of the boundary
conditions of the given configuration $\Phi$. 
All global $Z_2$ transformations must thus be considered equivalent \footnote{In contrast to the
transformations $T_\pm$ they are {\sl continuous} gauge transformations and 
thus do not change the physics.}. 
We conclude that it is two times more likely to generate antiperiodic boundary 
conditions out of a given 
periodic field configuration $\Phi$ than it is to maintain 
periodic boundary conditions if no a priori bias on possible Z$_2$ transformations 
along the $\tau$ direction exists. If the system is autonomous, which safely can be pressumed not too far
away from the phase boundary, there is 
no reason for such a bias to exist.      
\begin{figure}
\begin{center}
\leavevmode
\leavevmode
\vspace{5.3cm}
\includegraphics{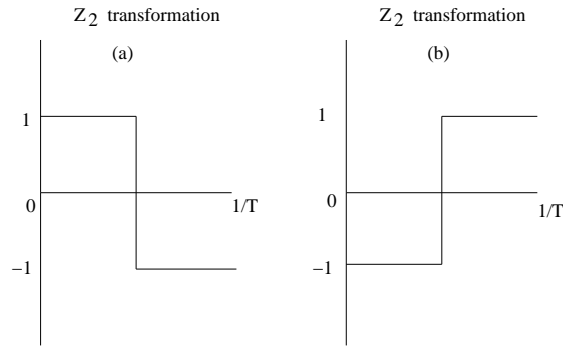}
\end{center}
\caption{Inequivalent, local Z$_2$ transformations along the Euclidean 
time coordinate at finite temperature.}      
\end{figure} 
At this point it is interesting to look back at Einstein's 
1925 paper \cite{Einstein1925}. In that paper an expression for 
the mean squared deviation of 
the energy of an electron gas was obtained. This expression 
separates into two parts. The first part was interpreted as a fluctuation 
term for distinguishable, Poisson distributed particles while 
the second term is due to 
indistinguishable particles or a wavefield a la de Broglie. 
Let us relate this our field theoretical result: 
The electron (and neutrino) ensemble is generated by 't Hooft-loop actions generating 
particles whose existence is based on the explicit occurrence of 
center-vortex loops or magnetic fluxes (see below). 
This ensemble is described by
antiperiodic boundary conditions (fermion). 
Whenever a particle is being created the condensate of center vortices 
it is immersed in proliferates the propagation 
of a longitudinal soundwave in the condensate. 
The associated degree of freedom is described by 
periodic boundary conditions (boson). Since a longitudinal 
soundwave is a scalar and fermions 
come in two spin orientations the statistical weighting is 1:2, 
respectively. 

We emphasize 
that the above discussion is a statistical one. 

{\sl Microscopic situation.} 
\begin{figure}
\begin{center}
\leavevmode
\leavevmode
\vspace{3.3cm}
\includegraphics{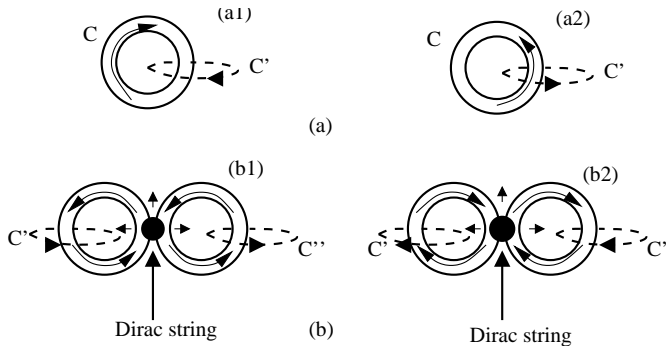}
\end{center}
\caption{The action of one 't Hooft loop (a) and the simultaneous action of 
two 't Hooft loops with the same sense of winding leading to a magnetic flux crossing (b). The latter 
is physically and topologically equivalent to an isolated magnetic monopole. 
The isotropic, magnetic flux associated with a flux crossing  
(indicated by small straight arrows), that is, the flux not confined to a 
center vortex is due to a Dirac string 
extending to infinity. Isotropic magnetic flux and Dirac string are present due to 
the theory SU(2)$_{\tiny{\mbox{CMB}}}$ not being in its confining phase. 
Distinguishing situations (a1), (a2) and (b1), (b2) is only possible 
if a direction is singled out in 3-space (a homogeneous magnetic field). 
The creation of the solitons (a) and (b) by 't Hooft loop action (indicated by the dashed curves 
$C^\prime$ and $C^{\prime\prime}$) takes place as each of the factor 
theories SU(2)$_{\tiny{e}}\times$SU(2)$_{\tiny{\mu}}\times$
SU(2)$_{\tiny{\tau}}$ undergoes the 1$^{\tiny\mbox{st}}$ order transition 
to the confining phase \cite{PRL1}.}      
\end{figure} 
In Fig.\,2 we indicate the particle content in the confining phase of one of the  
SU(2) gauge factors in Eq.\,(\ref{gs}). These particles are generated by 
the action of 't Hooft loops during the relaxation of the $\Phi$ expectation towards the minimum 
of the potential in Eq.\,(\ref{actN=2}). At the minimum of the potential the action of a 
't Hooft loop is energetically forbidden since 
the ground-state energy there is {\sl precisely} zero. 
A flux crossings is equivalent to 
an isolated magnetic monopole in the electric phase of SU(2) \cite{PRL1}: 
with regard to the intact U(1)$_{\tiny\mbox{SM}}$ gauge factor of the Standard 
model, which is the Cartan subgroup of SU(2)$_{\tiny\mbox{CMB}}$ \cite{PRL2}, 
it has one unit of electric charge.
If there were only a single SU(2) factor attributed to a 
realistic theory then a larger number of stable 
flux crossingss (or center vortex crossings) would be allowed 
for in one and the same soliton. However, 
since SU(2)$_{\tiny{\mbox{CMB}}}$ is not 
confining at the present temperature of the Universe, $T_{\tiny\mbox{CMB}}\sim 10^{-4}\,$eV, 
there is an asymptotic photon state \cite{PRL2} preventing the formation of stable 
solitons with multiple flux crossings.  The dynamical magnetic flux along a 
center vortex \footnote{
Imagine a dynamical chain of magnetic monopoles and antimonopoles 
with the monopoles moving to the right and the antimonopoles to the 
left \cite{Olejnik1997}.} in the gauge theory SU(2) is a dynamical electric flux 
with respect to U(1)$_{\tiny\mbox{SM}}$. With respect to the latter group it 
generates a magnetic field winding around the center vortex loop: 
the soliton is endowed with a {\sl magnetic moment} $\vec{\mu}_b$. The magnetic moment 
in soliton (b) is {\sl twice} 
as large as it is in soliton (a) (Fig.\,2)
\eqb
\label{magmoment}
|\vec{\mu}_b|=2\times |\vec{\mu}_a|\,.
\eqe
Since the two different flux orientations in each soliton are interpreted as intrinsic angular momentum 
1/2 (see below), which is the same for soliton (a) and (b), we conclude that the 
$g$ factor for a charged lepton is twice as high as that for a neutrino.      

The reader may easily 
convince himself that a soliton of the form (b) but with one of the 
center vortex loops rotated out of the
common plane is instable: 
The local interaction energy-density $\rho_{\tiny\mbox{int}}$ between the magnetic fields 
(w.r.t. U(1)$_{\tiny\mbox{SM}}$) $\vec{B}_1$ and $\vec{B}_2$ generated by the electric fluxes 
on either side of the isolated electric charge is 
given as 
\eqb
\label{magnets}
\rho_{\tiny\mbox{int}}= -\vec{B}_1\cdot \vec{B}_2\,.
\eqe
This is minimal for $\vec{B}_1\|\vec{B}_2$. 

Applying an external, static, and homogeneous magnetic field $\vec{B}$ 
(w.r.t. U(1)$_{\tiny\mbox{SM}}$) along an axis perpendicular 
to the center vortex loops in soliton (b) (Fig.\,2) and keeping the soliton fixed by means 
of a static electric field sensitive to the isolated electric charge
discriminates between the situations (b1) and (b2). The level splitting $\Delta E$
originating from the effective interaction of the magnetic moment 
$\vec{\mu}_b$ with the magnetic field $\vec{B}$
\eqb
\label{spins}
H_{\tiny\mbox{int}}^B=-\vec{\mu}_b\cdot \vec{B}
\eqe
is 
\eqb
\label{LS}
\Delta E=2|\vec{\mu}_b||\vec{B}|\,.
\eqe
In \cite{Hofmann2004} we have identified soliton (a) with 
the (Majorana) neutrino and soliton (b) with the charged lepton 
associated with one of the gauge-group 
factors in SU(2)$_{\tiny{e}}\times$SU(2)$_{\tiny{\mu}}
\times$SU(2)$_{\tiny{\tau}}$. 
The above situation is realized for a valence electron in an atom (soliton (b)), 
the level splitting of Eq.\,(\ref{LS} 
is known as the anomalous Zeeman effect. 
This effect was interpreted as an intrinsic angular
momentum $\frac{1}{2}$ of the electron 
in \cite{UhlenbeckGoudsmith1925,Compton1921}. 

{\sl Conclusion.} We have shown that the only stable solitons arising in the confining 
phases of each factor in $\mbox{SU(2)}_{\tiny{e}}\times \mbox{SU(2)}_{\tiny{\mu}}$ 
can be interpreted as
spin-1/2 fermions. Statistically, a fermions are inseparably linked 
to bosons. The latter are interpreted as 
longitudinal soundwaves propagating in the 
center-vortex condensate. So far we have 
not investigated how the weak interactions 
of the Standard Model can be 
understood in the solitonic framework presented here.    
 
{\sl Acknowledgment.} It is a pleasure to thank Holger 
Gies for useful conversations and Wolfgang Tvarusko for 
comments on the manuscript.

\bibliographystyle{prsty}

\end{document}